\documentclass[conference]{IEEEtran}

\usepackage[
  pass,
]{geometry}
\usepackage{multirow}
\usepackage{amsfonts}
\usepackage{epsfig}
\usepackage{amsmath}
\usepackage{amssymb}
\usepackage[acronym]{glossaries}
\newcommand\acro[2]{\newacronym{#1}{#1}{#2}}
\newcommand\acroAlwaysShort[1]{\newglossaryentry{#1}{type=\acronymtype, name={#1}, description={#1}, text={#1}, first={#1}, plural={#1s}, firstplural={#1s}}}
\newcommand\acroShortSurname[2]{\newglossaryentry{#1}{type=\acronymtype, name={#2}, description={#2}, text={#2}, first={#2}, plural={#2s}, firstplural={#2s}}}
\newcommand\ac[1]{\gls{#1}}
\newcommand\acp[1]{\glspl{#1}}
\newcommand\acs[1]{\glsname{#1}}

\acro{FoV}{Field of View}
\acro{ABR}{adaptive bit-rate}
\acro{DASH}{Dynamic Adaptive Streaming over HTTP}
\acro{CDN}{content delivery network}
\acro{CDF}{cumulative density function}
\acro{HMD}{Head-Mounted Display}
\acro{PDF}{probability density function}
\acro{RTP}{real-time protocol}
\acro{VR}{Virtual Reality}
\acro{QoE}{Quality of Experience}
\acro{VQM}{Video Quality Metric}
\acro{ILP}{integer linear program}
\acro{SRD}{Spatial Relationship Description}
\acro{HDTV}{high definition television}
\acro{UGC}{User-Generated Content}
\acro{MPD}{Media Presentation Description}
\acro{PSNR}{Peak Signal Noise to Ratio}
\acro{GPU}{graphics processing unit}
\acro{CPU}{central processing unit}
\acro{MS-SSIM}{Multiscale - Structural Similarity}
\acro{API}{Application Programming Interface}
\acro{QEC}{Quality Emphasis Center}
\acro{QER}{Quality Emphasized Region}
\newglossaryentry{RoI}{type=\acronymtype, name={RoI}, description={RoI}, text={RoI}, first={Region of Interest~(RoI)}, plural={RoI}, firstplural={Regions of Interest~(RoI)}}
\acro{VM}{Virtual Machine}
\acro{SVC}{Scalable Video Coding}
\acro{GOP}{Group of Picture}
\acro{PMU}{Performance Monitoring Unit}
\acro{LAN}{Local Area Network}
\acro{AI}{Artificial Intelligence}
\acro{3D}{Three Dimentional}
\acro{OS}{Operating System}
\newglossaryentry{fps}{type=\acronymtype, name={fps}, description={frame per second}, text={frame per second}, first={frame per second~(fps)}, plural={fps}, firstplural={frames per second~(fps)}}
\newglossaryentry{p}{type=\acronymtype, name={p}, description={p}, text={~pixel}, first={~pixel (p)}, plural={p}, firstplural={~pixels (p)}}
\newglossaryentry{s}{type=\acronymtype, name={s}, description={s},
text={second}, first={~second~(s)}, plural={s}, firstplural={~seconds~(s)}}
\acroAlwaysShort{TCP}
\acroAlwaysShort{HTTP}
\acroAlwaysShort{MPEG}
\acro{RTT}{Round-Trip Time}
\acro{AVC}{Advanced Video Coding}
\acro{HEVC}{High Efficiency Video Coding}
\acro{ISO}{International Organization for Standardization}
\newglossaryentry{ISOBMFF}{type=\acronymtype, name={ISOBMFF}, description={International Organization for Standardization base media file format}, text={International Organization for Standardization base media file format}, first={International Organization for Standardization base media file format ISO/IEC 14496-12 (ISO BMFF)}, plural={ISO BMFFs}, firstplural={International Organization for Standardization base media file formats (ISO BMFFs)}}
\acro{MTU}{Maximum Transmission Unit}
\acro{AQM}{Active Queue Management}
\acro{I}{intra-predicted}
\acro{P}{inter-predicted}
\acro{B}{bidirectional}
\acro{VQMT}{Video Quality Measurement Tool}
\acro{EPFL}{Ecole Polytechnique F\'{e}d\'{e}rale de Lausanne}
\acroShortSurname{YUV}{YUV}
\acroShortSurname{RGB}{R'G'B'}
\acro{MSE}{Mean Square Error}
\acro{CMSE}{Commulative Mean Square Error}
\usepackage[english]{babel}
\usepackage[numbers,sort]{natbib}
\renewcommand\citet[1]{\citeauthor{#1}~\cite{#1}} 
\usepackage{color}
\usepackage[dvipsnames,table]{xcolor}
\usepackage[linesnumbered,ruled,vlined,boxed,commentsnumbered]{algorithm2e}
\usepackage[noend]{algorithmic}
\usepackage{float}
\algsetup{linenosize=\tiny}
\usepackage[skip=5pt,font=bf,labelfont=bf]{caption}
\usepackage[caption=false]{subfig}
\usepackage{pgfplots}
\usepackage{booktabs}
\usepackage{listings}
\usetikzlibrary{shapes,positioning,3d,calc}
\usetikzlibrary{decorations,decorations.pathmorphing,backgrounds}
\usepgfplotslibrary{external}
\newcommand{\externaldirectory}{latex.out/}
\tikzexternaldisable
\usepackage{wrapfig}
\usepackage{enumitem}
\usepackage{url}
\usepackage{etoolbox}

\usepackage{siunitx}
\usepackage{mathtools}
\usepackage{stmaryrd}
\usepackage{mathrsfs}
\usepackage{amssymb}

\newcommand{\parag}[1]{\vspace{3pt}\noindent\textbf{#1}.\hspace{4pt}}

\usepackage{etoolbox}
\makeatletter
\patchcmd\@acf{\hskip\z@}{}{}{}
\patchcmd\@acf{\hskip\z@}{}{}{}
\makeatother

\usepackage{pgfplotstable}
\pgfplotsset{compat=newest}

\newbool{doubleBlinded}

\newbool{releaseMode}
\booltrue{releaseMode}  

\newbool{NotesActivated}
\ifbool{releaseMode}{\boolfalse{NotesActivated}}{\booltrue{NotesActivated}}

\newcommand{\noteGS}[1] {\ifbool{NotesActivated}{\color{red}\{\textbf{GS:}\textit{{#1}}\}\color{black}}{}}
\newcommand{\noteXC}[1] {\ifbool{NotesActivated}{\color{YellowOrange}\{\textbf{XC:}\textit{{#1}}\}\color{black}}{}}
\newcommand{\noteAD}[1] {\ifbool{NotesActivated}{\color{OliveGreen}\{\textbf{AD:}\textit{{#1}}\}\color{black}}{}}
\newcommand{\textWithColor}[2]{\ifbool{releaseMode}{#2}{\color{#1}#2\color{black}}}

\newcommand{\GS}[1] {\textWithColor{red}{#1}}

\newcommand\blfootnote[1]{%
  \begingroup
  \renewcommand\thefootnote{}\footnote{#1}%
  \addtocounter{footnote}{-1}%
  \endgroup
}

\newcommand\FoV[0]{viewport}

\setlist[itemize]{itemsep=-1pt,leftmargin=\parindent,topsep=2pt}
\setlist[enumerate]{itemsep=-1pt,leftmargin=\parindent,topsep=-2pt}
\setlist[description]{itemsep=-1pt,topsep=3pt}

\title{Viewport-Adaptive Navigable\\360-Degree Video Delivery}

\makeatletter

\author{\IEEEauthorblockN{Xavier Corbillon \hspace{10pt} Gwendal Simon}
\IEEEauthorblockA{IMT Atlantique, France\\
firstname.lastname@imt-atlantique.fr}
\and
\IEEEauthorblockN{Alisa Devlic}
\IEEEauthorblockA{Huawei Technologies, Sweden\\
alisa.devlic@huawei.com}
\and
\IEEEauthorblockN{Jacob Chakareski}
\IEEEauthorblockA{University Alabama, USA\\
jacob@ua.edu}}

\makeatother

\newbool{withColor}
\booltrue{withColor}

\ifbool{withColor} {
	\definecolor{color1}{HTML}{4e9a06}
	\definecolor{color2}{HTML}{c17d11}
	\definecolor{color3}{HTML}{cc0000}
	\definecolor{color4}{HTML}{204a87}
	\definecolor{color5}{HTML}{ad7fa8}

	\definecolor{titles}{HTML}{555753}

	\makeatletter
	\definecolor{beamer@tbbrown}{rgb}{0.43,0.31,0.28}
	\makeatother

	\definecolor{midqualityOri}{HTML}{e9b96e}
	\colorlet{midquality}{color2!60!white}
	\colorlet{fullquality}{black!20!color2}
}{
	\colorlet{color1}{black!25}
	\colorlet{color2}{black!50}
	\colorlet{color3}{black!75}
	\colorlet{color4}{black!100}
	\colorlet{color5}{gray!50}

	\colorlet{titles}{gray!80}
	\colorlet{midquality}{gray!30}
	\colorlet{fullquality}{gray!70}
}

\tikzset{
	spherical/.pic={
		\shade[ball color=midquality,opacity=0.60] (0,0) circle (#1 pt);
		\draw (-0.0352778*#1,0) arc (180:360:#1 pt and 0.5*#1 pt);
	    \draw[densely dashed] (-0.0352778*#1,0) arc (180:0:#1 pt and 0.5*#1 pt);
   	    \draw (0,0.0352778*#1) arc (90:270:0.5*#1 pt and #1 pt);
   	    \draw[densely dashed] (0,0.0352778*#1) arc (90:-90:0.5*#1 pt and #1 pt);
     	\draw (0,0) circle (#1 pt);
    }
}

\tikzset{
	pics/equirectangular/.style n args={3}{
		code={
			\draw[fill=midquality] (#2*0.00881945*#1-3*0.00881945*#1,#3*0.004961*#1-3*0.004961*#1)
			rectangle
			(#2*0.00881945*#1+4*0.00881945*#1,#3*0.004961*#1+4*0.004961*#1);

			\draw[fill=fullquality] (#2*0.00881945*#1-0.00881945*#1,#3*0.004961*#1-0.004961*#1)
			rectangle
			(#2*0.00881945*#1+2*0.00881945*#1,#3*0.004961*#1+2*0.004961*#1);

			\foreach \i in {-4,-3,-2,-1,0,1,2,3}{
				\foreach \j in {-4,-3,-2,-1,0,1,2,3}{
					\draw (\i*0.00881945*#1,\j*0.004961*#1) rectangle (\i*0.00881945*#1+0.00881945*#1,\j*0.004961*#1+0.004961*#1);
					}
				}
		}
	}
}

\tikzset{
	cubemap/.pic={
		\draw[fill=midquality] (-0.0352778*#1,-0.006615*#1) rectangle (-0.017639*#1,0.006615*#1);
		\draw[fill=fullquality] (-0.017639*#1,-0.006615*#1) rectangle (0,0.006615*#1);
		\draw[fill=midquality] (0,-0.006615*#1) rectangle (0.017639*#1,0.006615*#1);
		\draw[fill=white] (0.017639*#1,-0.006615*#1) rectangle (0.0352778*#1,0.006615*#1);
		\draw[fill=midquality] (-0.017639*#1,0.006615*#1) rectangle (0,0.019844*#1);
		\draw[fill=midquality] (-0.017639*#1,-0.006615*#1) rectangle (0,-0.019844*#1);
	}
}

\tikzset{
	pyramid/.pic={
		\draw[fill=fullquality] (-0.011759*#1,-0.006615*#1) rectangle (0.011759*#1,0.006615*#1);
		\draw[fill=midquality] (-0.011759*#1,0.006615*#1)
			 -- (0.011759*#1,0.006615*#1)
			 -- (0,0.019844*#1)
			 -- cycle;
		\draw[fill=midquality] (-0.011759*#1,-0.006615*#1)
			 -- (0.011759*#1,-0.006615*#1)
			 -- (0,-0.019844*#1)
			 -- cycle;
		\draw[fill=midquality] (-0.011759*#1,-0.006615*#1)
			 -- (-0.011759*#1,0.006615*#1)
			 -- (-0.0352778*#1,0)
			 -- cycle;
		\draw[fill=midquality] (0.011759*#1,-0.006615*#1)
			 -- (0.011759*#1,0.006615*#1)
			 -- (0.0352778*#1,0)
			 -- cycle;
	}
}

\tikzset{
	pics/losange/.style n args={3}{
		code={
			\draw[fill=#2, rotate around={#3:(0,0)}]
				(0,0)
				-- (#1, 0.75*#1)
				-- (2*#1, 0)
				-- (#1, -0.75*#1)
				--	cycle;
		}
	}
}

\tikzset{
	dodecahedron/.pic={
		\def\xshi{0}
		\pic at(\xshi,0) {losange={#1}{white}{0}};
		\pic at(\xshi,0) {losange={#1}{white}{287}};
		\pic at(\xshi+2*#1,0) {losange={#1}{midquality}{254}};
		\pic at(\xshi+1.46*#1,-1.93*#1) {losange={#1}{midquality}{0}};

		\def\xshi{2.89*#1}
		\pic at(\xshi,0) {losange={#1}{fullquality}{0}};
		\pic at(\xshi,0) {losange={#1}{fullquality}{287}};
		\pic at(\xshi+2*#1,0) {losange={#1}{midquality}{254}};
		\pic at(\xshi+1.46*#1,-1.93*#1) {losange={#1}{white}{0}};

		\def\xshi{-2.89*#1}
		\pic at(\xshi,0) {losange={#1}{midquality}{0}};
		\pic at(\xshi,0) {losange={#1}{midquality}{287}};
		\pic at(\xshi+2*#1,0) {losange={#1}{midquality}{254}};
		\pic at(\xshi+1.46*#1,-1.93*#1) {losange={#1}{white}{0}};
	}
}

\begin{document}

\maketitle

\begin{abstract}
The delivery and display of 360-degree videos on Head-Mounted Displays (HMDs) presents 
many technical challenges. 360-degree videos are ultra high resolution spherical videos, which 
contain an omnidirectional view of the scene. However only a portion of this scene is displayed 
on the HMD. Moreover, HMD need to respond in 10 ms to head movements, which prevents the 
server to send only the displayed video part based on client feedback. To reduce the bandwidth 
waste, while still providing an immersive experience, a viewport-adaptive 360-degree video 
streaming system is proposed. The server prepares multiple video representations, which differ not 
only by their bit-rate, but also by the qualities of different scene regions. The client chooses a 
representation for the next segment such that its bit-rate fits the available throughput and a full 
quality region matches its viewing. We investigate the impact of various spherical-to-plane 
projections and quality arrangements on the video quality displayed to the user, showing that the 
cube map layout offers the best quality for the given bit-rate budget. An evaluation with a dataset 
of users navigating 360-degree videos demonstrates that segments need to be short enough to 
enable frequent view switches.
\end{abstract}
\blfootnote{The work of J. Chakareski was supported by NSF award CCF-1528030}

\glsresetall

\section{Introduction}
\label{sec:introduction}

The popularity of navigable 360-degree video systems has grown with
the advent of omnidirectional capturing systems
and interactive displaying systems, like \acp{HMD}. However, to
deliver 360-degree video content on the Internet, the content
providers have to deal with a problem of bandwidth waste: What is
displayed on the device, which is indifferently called
\textit{\ac{FoV}} or \textit{viewport}, is only a fraction of what is
downloaded, which is an omnidirectional view of the scene.
This bandwidth waste is the price to pay for interactivity. To prevent
\emph{simulator sickness}~\cite{moss2011characteristics} and to
provide good \ac{QoE}, the vendors of \acp{HMD} recommend that the
enabling multimedia systems react to head movements as fast as the
\ac{HMD} refresh rate.
Since the refresh rate of state-of-the-art \acp{HMD} is
\SI{120}{\hertz},
the whole
system should react in less than \SI{10}{ms}. This delay constraint
prevents the implementation of traditional delivery architectures
where the client notifies a server about changes and awaits for the
reception of content adjusted at the server. Instead, in the current
\ac{VR} video delivery systems, the server sends the full $360$-degree
stream, from which the \ac{HMD} extracts the viewport in real time,
according to the user head movements. Therefore, the majority of the
delivered video stream data are not used.

Let us provide some numbers to illustrate this problem. The viewport
is defined by a device-specific viewing angle (typically
$120$ degrees), which delimits horizontally the scene from the head direction center, called \FoV{} center. To ensure a
good immersion, the pixel resolution of the displayed viewport is high,
typically $4$K ($3840\times2160$). So the
resolution of the full $360$-degree video is at least $12$K
($11520\times6480$). In addition, the immersion requires a video frame
rate on the order of the \ac{HMD} refresh rate, so typically around
\SI[mode=text]{100}{\acp{fps}}. Overall, high-quality $360$-degree
videos combine both a very large resolution (up to $12$K) and a very
high frame rate (up to \SI[mode=text]{100}{\acp{fps}}). To compare,
the bit-rate of 8K videos at \SI[mode=text]{60}{\acp{fps}} encoded
using \ac{HEVC} is around \SI{100}{Mbps}~\cite{7398367}.

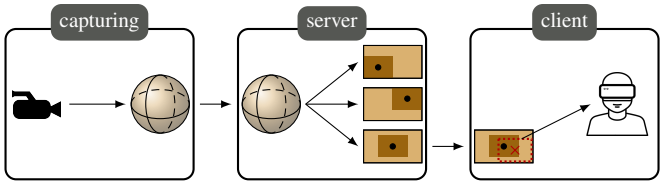
\begin{figure}
   \centering
\tikzsetnextfilename{delivery_chain}
\begin{tikzpicture}

\tikzset{
     element/.style={
     	rounded corners,
     	rectangle,
  	 	thick,
  	 	draw=black,
  	 	minimum height=2cm,minimum width=2.5cm
     }
}

\tikzset{
	elementtitle/.style={
		rectangle,
		rounded corners,
		fill=titles,
		font=\footnotesize,
		text=white,
		anchor=north
	}
}

\tikzset{
	pics/equirec/.style n args={3}{
		code={
			\draw[fill=midquality] (-0.0352778*#1, -0.019844*#1) rectangle (0.0352778*#1, 0.019844*#1);
			\draw[draw=none,fill=fullquality] (0.0088194*#2*#1-2*0.0088194*#1, 0.0066147*#3*#1 - 2*0.0066147*#1) rectangle (0.0088194*#2*#1 + 2*0.0088194*#1, 0.0066147*#3*#1 + 2*0.0066147*#1);
			\draw[draw,fill=none] (-0.0352778*#1, -0.019844*#1) rectangle (0.0352778*#1, 0.019844*#1);
			\draw[color=black,fill=black] (0.0088194*#2*#1, 0.0066147*#3*#1) circle (1pt);
		}
	}
}

\def\convCmPt{0.0352778}
\def\convCmPtRec{0.019844}
\def\convCmPtRecThird{0.0066147}
\def\convCmPtFourth{0.0088194}

\tikzset{cross/.style={cross out, draw,
         minimum size=2*(#1-\pgflinewidth),
         inner sep=0pt, outer sep=0pt}}

\tikzset{
	fov/.pic ={
		\draw[densely dotted, thick, red!70!black] (-0.07,0.10) rectangle (0.37,-0.20);
		\draw (0.15,-0.05) node[cross=2pt,red!70!black] {};
	}
}

\tikzset{
	vr/.pic = {
		\draw[rounded corners] (-0.0352778*#1, -0.019844*#1) rectangle (0.0352778*#1, 0.019844*#1);
		\draw[rounded corners, thick] (-0.032*#1, -0.019844*#1) rectangle (0.032*#1, 0.016*#1);
		\node[font=\scriptsize,rectangle,red, draw=red, thick,
					densely dotted, anchor=east, inner sep=2pt,
					yshift=-1pt, xshift=-1pt] at (0,0) {L};
		\node[font=\scriptsize,rectangle,red, draw=red, thick,
					densely dotted, anchor=west, inner sep=2pt,
					yshift=-1pt,xshift=0.5pt] at (0,0) {R};
	}
}

\def\ecartElement{16pt}
\def\sizeSphere{11}
\def\ecartObjet{2}
\def\ecartYVersions{16pt}

\node[element] (0,0) (capturing) {};
\node[elementtitle, above=-5pt of capturing] {capturing};
\draw ([xshift=-\sizeSphere - \ecartObjet pt]capturing.east) pic {spherical=\sizeSphere};
\pgfdeclareimage[width=18 pt]{camera}{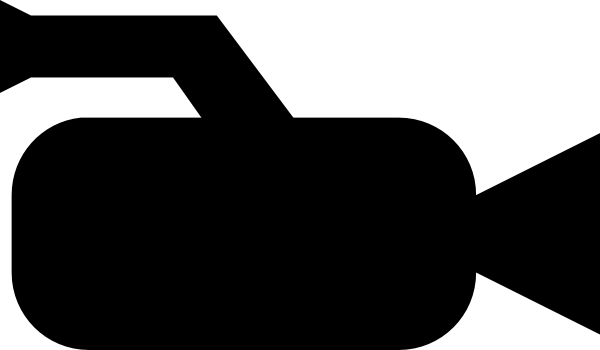}
\node at ([xshift=21 pt]capturing.west) (camera1)
    {\pgfbox[right,center]{\pgfuseimage{camera}}};

\draw[-latex] (camera1.east) to ([xshift=-2*\sizeSphere - 2*\ecartObjet pt]capturing.east);

\node[element,right=\ecartElement of capturing] (server) {};
\node[elementtitle, above=-5pt of server] {\vphantom{pt}server};
\draw ([xshift=\sizeSphere + \ecartObjet pt]server.west) pic {spherical=\sizeSphere};
\draw ([xshift=-\sizeSphere - \ecartObjet pt]server.east) pic {equirec={\sizeSphere}{2}{1}};
\draw ([xshift=-\sizeSphere - \ecartObjet pt, yshift=\ecartYVersions]server.east) pic {equirec={\sizeSphere}{-2}{-1}};
\draw ([xshift=-\sizeSphere - \ecartObjet pt, yshift=-\ecartYVersions]server.east) pic {equirec={\sizeSphere}{0}{0}};

\draw[-latex] ([xshift=2*\sizeSphere + 2*\ecartObjet pt]server.west) to ([xshift=-2*\sizeSphere - 2*\ecartObjet pt]server.east);
\draw[-latex] ([xshift=2*\sizeSphere + 2*\ecartObjet pt]server.west) to ([xshift=-2*\sizeSphere - 2*\ecartObjet pt, yshift=\ecartYVersions]server.east);
\draw[-latex] ([xshift=2*\sizeSphere + 2*\ecartObjet pt]server.west) to ([xshift=-2*\sizeSphere - 2*\ecartObjet pt, yshift=-\ecartYVersions]server.east);

\draw[-latex] ([xshift=2pt]capturing.east) to ([xshift=-2pt]server.west);

\node[element,right=\ecartElement of server] (client) {};
\node[elementtitle, above=-5pt of client] {\vphantom{pt}client};
\draw([xshift=\sizeSphere + \ecartObjet pt, yshift=-\ecartYVersions]client.west) pic {equirec={\sizeSphere}{0}{0}};
\pic[local bounding box=thisfov] at ([xshift=\sizeSphere + \ecartObjet pt, yshift=-\ecartYVersions]client.west) {fov};
%

\pgfdeclareimage[width=24 pt]{vrheadset}{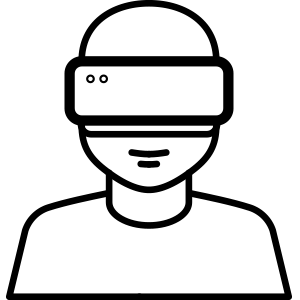}
\node at ([xshift=-28 pt]client.east) (headset)
    {\pgfbox[left,center]{\pgfuseimage{vrheadset}}};

%

\draw[-latex]
	(thisfov.60) to
		([xshift=-2*\sizeSphere - 2*\ecartObjet pt]client.east);

\draw[-latex] ([xshift=2pt, yshift=-\ecartYVersions]server.east) to ([xshift=-2pt, yshift=-\ecartYVersions]client.west);

\end{tikzpicture}
   \caption{Viewport-adaptive 360-degree video delivery system: The server
   offers video representations for three \acp{QER}. The dark \ifbool{withColor}{brown}{gray} is the part of the video encoded at high quality, the light
   \ifbool{withColor}{brown}{gray} the low quality. The viewport is the dotted red rectangle, the \FoV{} center the
   cross}
   \label{fig:deliverychain}
\end{figure}

We propose in this paper a solution where, following the same
principles as in rate-adaptive streaming technologies, the server
offers multiple \emph{representations} of the same $360$-degree video.
But instead of offering representations that only differ by their
bit-rate, the server offers here representations that differ by having
a \ac{QER}: a region of the video with a better quality than the remaining
of the video.
Our proposal is a
\emph{viewport-adaptive streaming system} and is depicted in
Figure~\ref{fig:deliverychain}. The \ac{QER} of each video representation is characterized
by a \emph{\ac{QEC}}, which is the center of the \ac{QER} and represents a given viewing position in the
spherical video. Around the \ac{QEC}, the quality of the video is
maximum, while it is lower for video parts that are far from the
\ac{QEC}. Similarly as in \ac{DASH}, the video is cut into segments
and the client periodically runs an \emph{adaptive algorithm} to
select a representation for the next segment. In a
viewport-adaptive system, clients select the representation
such that the bit-rate fits their receiving
bandwidth and the \ac{QEC} is closest to their \FoV{} center.

This viewport-adaptive $360$-degree streaming system has three
advantages: $(i)$ the bit-rate of the delivered video is lower than
the original full-quality video because video parts distant from the
\ac{QEC} are encoded at low quality. $(ii)$ When the end-user does not
move, the viewport is extracted from the highest quality part of the
spherical video. And $(iii)$ when the head of the end-user moves, the
device can still extract a viewport because it has the full
spherical video. If the new \FoV{} center is far from the \ac{QEC}
of the received video representation, the quality of the
extracted viewport is lower but this degradation holds only until the
selection of another representation with a closer \ac{QEC}.

The remainder of the paper is organized as follows. First, we
present our viewport-adaptive streaming
system, and
we show how it can be integrated into
the \ac{MPEG} \ac{DASH}-VR standard. Our proposal is thus a contribution
to the \ac{VR} group that \ac{MPEG} launched in May
$2016$~\cite{mpeg-vr}. Second, we address the choice of the geometric
layout into which the spherical video is projected for
encoding. We evaluate several video quality arrangements for a given
geometric layout and show that the cube map layout with full quality around the \ac{QEC} and \SI{25}{\percent} of this quality in the remaining faces offers the best quality of the extracted viewport.
Third, we study the required video segment length for
viewport-adaptive streaming. Based on a dataset of real users
navigating $360$-degree videos, we show that head movements occur over
short time periods, hence the streaming video segments have to be
short enough to enable frequent \ac{QER} switches. Fourth, we
examine the impact of the number of \acp{QER} on the viewport quality
and we show that a small number of (spatially-distributed over the sphere)
\acp{QER} suffices to get high viewport quality.
Finally, we introduce a tool (released as open source), which creates video representations for the proposed viewport-adaptive streaming system.
The tool is highly configurable: from a given
$360$-degree video, it allows any arrangement of video quality for a
given geometric layout, and it extracts the viewport from any \FoV{} center.
This tool thus provides the main software module
for the implementation of viewport-adaptive streaming of navigable
$360$-degree videos.

\section{Background and Related Work}
\label{sec:related}

We introduce the necessary geometric concepts for spherical
videos, and discuss prospective architecture
proposals for navigable $360$-degree video delivery.

\subsection{Geometric Layouts for 360-degree Videos}

\begin{figure}[t]
\centering
\tikzsetnextfilename{2d_layout_projections}
\begin{tikzpicture}
\def\sizeSphere{20}\def\ecartY{-1.2}\def\ecartX{6}

\pic [local bounding box=spher]  at (0,0) {spherical=15};

\pic [local bounding box=equi] at (-3,\ecartY) {equirectangular={\sizeSphere}{-1}{0}};

\pic [local bounding box=cubemap] at (-1,\ecartY) {cubemap=\sizeSphere};

\pic [local bounding box=pyra] at (1,\ecartY) {pyramid=\sizeSphere};


\def\unitused{0.22}

\pic [local bounding box=dodeca] at (3,0.88*\ecartY) {dodecahedron=\unitused};

\draw[-latex] (spher.180) -| (equi);
\draw[-latex] (spher.200) -| (cubemap);
\draw[-latex] (spher.340) -| (pyra);
\draw[-latex] (spher) -| (dodeca);

\node[font=\scriptsize,anchor=north] at (equi.south) {equirectangular};
\node[font=\scriptsize,anchor=north] at (cubemap.south) {cube map};
\node[font=\scriptsize,anchor=north] at (pyra.south) {pyramid};
\node[font=\scriptsize,anchor=north] at (dodeca.south) {\vphantom{y}dodecahedron};

\end{tikzpicture}
\caption{Projections into four geometric layouts}\label{fig:mapping}
\end{figure}
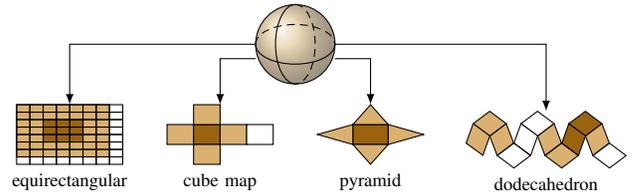

A $360$-degree video is captured in every direction from a unique point,
so it is essentially a \emph{spherical} video. Since current video encoders
operate on a two-dimensional rectangular image, a key step of the
encoding chain is to project the spherical video onto a planar
surface. The projection of a sphere onto a plane (known as mapping)
has been studied for centuries. In this paper, we consider the four
projections that are the most discussed for $360$-degree video
encoding~\cite{yu_framework_2015}. These layouts are depicted in
Figure~\ref{fig:mapping}.
From the images that are projected on an \textit{equirectangular}
panorama, a \textit{cube map}, and a \textit{rhombic dodecahedron}, it
is possible to generate a viewport for any position and angle in the
sphere without any information loss~\cite{Ng2005, fu_rhombic_2009}.
However, some pixels are over-sampled (a pixel on the sphere is
projected to a pair of pixels in the projected image). This is
typically the case for the sphere pole when projected on the
equirectangular panorama. This over-sampling degrades the performance
of traditional video encoders~\cite{
yu_framework_2015}. On the contrary, the projection into a pyramid
layout causes under-sampling: some pairs of pixels on the sphere
are merged into a
single pixel in the projected image by interpolating their color
values. This under-sampling cause distortion and information loss in
some extracted \FoV{}s. Previous work regarding projection of spherical
videos into different geometric layouts focuses on enabling efficient
implementation of signal processing
functions~\cite{kazhdan_metric-aware_2010} and improving the video
encoding~\cite{tosic_low_2009}.

\parag{Our contributions}We propose to leverage the geometric
structure of the layouts to implement a video encoding based on
\ac{QER}. Each geometric layout is characterized by a number of
\emph{faces} (\textit{e.g.}, 6 for the cube map, 12 for the
dodecahedron) and a given \emph{central point} (which corresponds to a
position on the sphere).
From the given central point and layout, our idea is to encode
the front face in full quality while the quality of other faces is
reduced.
To our knowledge, such idea has not been studied yet.
Another originality of our work is that we measure \ac{QoE} by
measuring the quality of several extracted viewports instead of
the full projected video.

\subsection{Personalized Viewport-Only Streaming}

An intuitive idea to address the problem of resource waste due to the
delivery of non-displayed video data is to stream only the part of the video that
corresponds to the viewport. This solution however does not enable
fast navigation within the $360$-degree video: When the client moves the
head, the \FoV{} center changes, requiring a new viewport to be
immediately displayed. Since the device has no knowledge about other
parts of the spherical video, it has to notify the server about the
head movement and wait for the reception of the newly adjusted viewport.
As seen in other interactive multimedia
systems~\cite{ChoyWSR14}, this solution cannot meet the \SI{10}{ms} latency
requirement in the standard Internet, even with the assistance of
\ac{CDN}. In addition, this solution requires the server to extract a
part of the video (thus to spend computing resources) for each client
connection.

\parag{Our contributions}In our system, the server always delivers
the full video, but it has different versions of this video depending
on the \ac{QER} (characterized by its \ac{QEC}). The client device
selects the right representation and extracts the viewport. The
storage requirements at the server side increase but all the
processing is done at the client side (representation selection and
viewport extraction). This idea matches the adaptive delivery
solutions that content providers have recently adopted (\textit{e.g.}~\ac{DASH}),
trading client-personalized delivery for
simple server-side management operation.

\subsection{Tiling for Adaptive Video Streaming}
To deal with the cases of end-users consuming only a fraction of the
video (navigable
panorama~\cite{sanchez_compressed_2015,wang_mixing_2014,gaddam_tiling_2015}
and large-resolution video~\cite{jean16mmsys}), the most studied delivery solution
leverages the concept
of \emph{tiling}.
The idea is to
spatially cut a video into independent tiles. The server offers
multiple video representations of each tile; the client periodically
selects a representation for each tile and it has to reconstruct the
full video from these tiles before the viewport extraction.
In a short paper,~\citet{ochi_live_2015} have sketched a
tile-based streaming system for $360$-degree videos. In their proposal,
the spherical video is mapped onto an \emph{equirectangular} video,
which is cut into $8\!\times\! 8$ \emph{tiles}.
More recently,~\citet{vishyArxiv} proposed a \emph{hexaface sphere}-based tiling
of a $360$-degree video to take into account projection distortion. They also present
an approach to describe the tiles with \ac{MPEG} \ac{DASH} \ac{SRD}
formatting principles. \citet{allthings} also propose the delivery of tiles based on
a prediction of the head movements. \citet{zare2016hevc} evaluate the impact of
different tiling scheme on the compression efficiency and on the transmission
bit-rate saving.

A tile-based adaptive streaming system provides the same features as
our proposed system regarding navigability (the clients get the full
video), bandwidth waste reduction (the video at low quality for
non-\FoV{} part) and \ac{QoE} maintenance (the downloaded video is
at full quality near the \FoV{} center). It has however several
critical weaknesses. First, the client has to first reconstruct the
video from independent tiles before the viewport extraction can take
place, which requires energy and time spent for each video frame.
Second, the more tiles there are, the less efficient the video
encoding is due to the tile
independence~\cite{sanchez_compressed_2015}. Third, the management at
the server is heavier because the number of files is larger. For
example, a typical $8\times8$ tiling offered at six quality levels
contributes to having $384$~independent files for each video segment,
and this results in larger \ac{MPD} files (or manifest
files). Finally, the management at the client side is heavier. 
For each tile, the client should run a representation selection
process and manage a specific network connection with the server.

\parag{Our contributions}In our system, the server
prepares $n$ \ac{QER}-based videos, each of them being a
pre-processed set of tile representations. Each \ac{QER}-based video is then
encoded at $k$ \emph{global} quality levels.
The main advantages include
an easier management for the server (fewer files hence a smaller
\ac{MPD} file), a simpler selection process for the client (by a
distance computation), and no need for re-constructing the video before
the viewport extraction.

\subsection{QER-Based Streaming}

A $360$-degree video provider (Facebook) has recently
released detailed the implementation of its delivery
platform~\cite{facebook}.
The spherical video is projected onto a pyramid layout from up to $30$ central
points to generate a set of video representations.
Since the
front face of pyramid projection has a better image quality than the other faces, the
system is in essence similar to our concept of \ac{QER}. The end-users
periodically select one of the representations
based on their \FoV{} center. This implementation
corroborates that, from an industrial perspective, the extra-cost of
generating and storing multiple \ac{QER}-based representations of the
same video is compensated by bandwidth savings and
enhanced system usability.
However, as seen in Section~\ref{sec:settings}, the pyramid projection is not
the best regarding the viewport quality. Moreover, the system
uses the same video quality on each face, which is less
efficient than our proposal. Finally, the impact of the video encoding on the solution
is not given.

\citet{lee2011efficient} studied in another context the coding of a regular
video with a \ac{QER}. The \ac{QER} is generated near the area that is the most likely
to attract gazes. They do not propose to generate different representations with different
\acp{QER}.

\parag{Our contributions}Our approach is based on the same
idea of offering multiple \ac{QER}-based video representations.
However, we provide a complete study of our system with the additional
distinction of having varying quality across the geometrical layout. Moreover,
our study includes an evaluation of several
geometric layouts, an analysis of the best segment duration, an
analysis of the best number of \acp{QER},
and a step towards integration into MPEG \ac{DASH}.


\section{System architecture}

This section describes the system architecture of the proposed
navigable $360$-degree video delivery framework.

\begin{figure}
   \centering
\tikzsetnextfilename{new_delivery}
\begin{tikzpicture}

\tikzset{
     element/.style={
     	rounded corners,
     	rectangle,
  	 	thick,
  	 	draw=black,
  	 	minimum height=2cm,minimum width=2.5cm
     }
}

\tikzset{
	elementtitle/.style={
		rectangle,
		rounded corners,
		fill=titles,
		font=\footnotesize,
		text=white,
		anchor=north
	}
}

\tikzset{
	pics/equirec/.style n args={3}{
		code={
			\draw[fill=midquality] (-0.0352778*#1, -0.019844*#1) rectangle (0.0352778*#1, 0.019844*#1);
			\draw[draw=none,fill=fullquality] (0.0088194*#2*#1-2*0.0088194*#1, 0.0066147*#3*#1 - 2*0.0066147*#1) rectangle (0.0088194*#2*#1 + 2*0.0088194*#1, 0.0066147*#3*#1 + 2*0.0066147*#1);
			\draw[draw,fill=none] (-0.0352778*#1, -0.019844*#1) rectangle (0.0352778*#1, 0.019844*#1);
			\draw[color=black,fill=black] (0.0088194*#2*#1, 0.0066147*#3*#1) circle (1pt);
		}
	}
}

\tikzset{
	emptyEquirec/.pic={
		\draw[fill=midquality] (-0.0352778*#1, -0.019844*#1) rectangle (0.0352778*#1, 0.019844*#1);
	}
}

\def\convCmPt{0.0352778}
\def\convCmPtRec{0.019844}
\def\convCmPtRecThird{0.0066147}
\def\convCmPtFourth{0.0088194}

\tikzset{cross/.style={cross out, draw,
         minimum size=2*(#1-\pgflinewidth),
         inner sep=0pt, outer sep=0pt}}

\tikzset{
	pics/fov/.style n args={2}{
		code = {
			\draw (-0.5*#1, -0.28*#1) rectangle (0.5*#1, 0.28*#1);
			\draw[densely dotted, thick, red!70!black] (-0.45*#1 + #2*#1*0.24, 0.15*#1)
				 rectangle
			 	(-0.4*#1 + #2*#1*0.24 + 0.4*#1, -0.15*#1);
			\draw (-0.4*#1 + #2*#1*0.24 + 0.2*#1, 0.05*#1) node[cross=2pt,red!70!black] {};
	   }
	}
}

\tikzset{
	vr/.pic = {
		\draw[rounded corners] (-0.0352778*#1, -0.019844*#1) rectangle (0.0352778*#1, 0.019844*#1);
		\draw[rounded corners, thick] (-0.032*#1, -0.019844*#1) rectangle (0.032*#1, 0.016*#1);
		\node[font=\scriptsize,rectangle,red, draw=red, thick,
					densely dotted, anchor=east, inner sep=2pt,
					yshift=-1pt, xshift=-1pt] at (0,0) {L};
		\node[font=\scriptsize,rectangle,red, draw=red, thick,
					densely dotted, anchor=west, inner sep=2pt,
					yshift=-1pt,xshift=0.5pt] at (0,0) {R};
	}
}

\tikzset{
	pics/threerep/.style n args={5}{
		code={
			\pic[local bounding box=bigA] at (0,0) {equirec={\sizeBig}{#4}{#5}};
			\ifnum#1>0
		    	\node[font=\scriptsize, anchor=east, inner sep=0pt]
		    		at (bigA.west) (legbwhi) {high};
		    \fi
		    \ifnum#3>0
				\node[font=\scriptsize,anchor=south] at (bigA.north) {$s_{#3}$};
			\fi
			\pic[local bounding box=lowA] at ([yshift=\ecartLow]bigA.south)
				{equirec={\sizeLow}{#4}{#5}};
			\ifnum#1>0
				\node[font=\scriptsize, inner sep=0pt, anchor=west]
					at (legbwhi.west |- lowA) {\vphantom{g}low};
			\fi
			\ifnum#2>0
				\node[font=\footnotesize,anchor=east] at (legbwhi.south west) {QER$_#2$};
			\fi
		}
	}
}

\tikzset{
	leftVR/.pic = {
		\draw plot [smooth] coordinates {(0,0)
			(0.05*#1,0.2*#1)
			(0.25*#1,0.2*#1)
			(0.3*#1,0)};
		\draw (0,0) -- (0.3*#1,0);
		\draw[fill=white] (0.15*#1, 0.4*#1) ellipse (0.15*#1 cm and 0.2*#1 cm);
		\draw[rounded corners=0.3*#1, fill=black] (-0.03*#1, 0.55*#1) rectangle
			(0.1*#1, 0.35*#1);
		\begin{scope}
			\clip (0.15*#1, 0.4*#1) ellipse (0.15*#1 cm and 0.2*#1 cm);
			\draw[fill=black] (0.1*#1, 0.5*#1) rectangle (0.3*#1, 0.46*#1);
			\draw plot [smooth] coordinates {(-0.01*#1,0.3*#1)
			 (0.04*#1,0.27*#1)
			 (0.09*#1,0.3*#1)};
		\end{scope}
	}
}

\tikzset{
	rightVR/.pic = {
		\draw plot [smooth] coordinates {(0,0)
			(0.05*#1,0.2*#1)
			(0.25*#1,0.2*#1)
			(0.3*#1,0)};
		\draw (0,0) -- (0.3*#1,0);
		\draw[fill=white] (0.15*#1, 0.4*#1) ellipse (0.15*#1 cm and 0.2*#1 cm);
		\draw[rounded corners=0.3*#1, fill=black] (0.33*#1, 0.55*#1) rectangle
			(0.2*#1, 0.35*#1);
		\begin{scope}
			\clip (0.15*#1, 0.4*#1) ellipse (0.15*#1 cm and 0.2*#1 cm);
			\draw[fill=black] (0.2*#1, 0.5*#1) rectangle (0, 0.46*#1);
			\draw plot [smooth] coordinates {(0.22*#1,0.3*#1)
			 (0.27*#1,0.27*#1)
			 (0.32*#1,0.3*#1)};
		\end{scope}
	}
}

\tikzset{
	frontVR/.pic = {
		\draw plot [smooth] coordinates {(0,0)
			(0.08*#1,0.2*#1)
			(0.42*#1,0.2*#1)
			(0.5*#1,0)};
		\draw (0,0) -- (0.5*#1,0);
		\draw[fill=white] (0.25*#1, 0.4*#1) ellipse (0.15*#1 cm and 0.2*#1 cm);
		\draw plot [smooth] coordinates {(0.18*#1,0.3*#1)
			 (0.25*#1,0.27*#1)
			 (0.32*#1,0.3*#1)};
		\draw[rounded corners=0.5*#1, fill=black] (0.07*#1, 0.55*#1) rectangle
			(0.43*#1, 0.35*#1);
	}
}

\def\sizeBig{11}
\def\sizeMed{9}
\def\ecartMed{-6 pt}
\def\sizeLow{7}
\def\ecartLow{-6 pt}
\def\ecartInterThree{-0.52}

\pic[local bounding box=leftup] at (0,0) {threerep={1}{1}{1}{2}{1}};
\pic[anchor=east, local bounding box=leftmid] at ($(-0.4,\ecartInterThree)+(leftup.east |- leftup.south)$)
	  {threerep={1}{2}{0}{-2}{-1}};
\pic[anchor=east, local bounding box=leftdown] at ($(-0.4,\ecartInterThree)+(leftmid.east |- leftmid.south)$)
	  {threerep={1}{3}{0}{0}{0}};

\pic[local bounding box=centerup] at (1,0) {threerep={0}{0}{2}{2}{1}};
\pic[anchor=east, local bounding box=centermid] at ($(-0.4,\ecartInterThree)+(centerup.east |- centerup.south)$)
	  {threerep={0}{0}{0}{-2}{-1}};
\pic[anchor=east, local bounding box=centerdown] at ($(-0.4,\ecartInterThree)+(centermid.east |- centermid.south)$)
	  {threerep={0}{0}{0}{0}{0}};

\pic[local bounding box=rightup] at (2,0) {threerep={0}{0}{3}{2}{1}};
\pic[anchor=east, local bounding box=rightmid] at ($(-0.4,\ecartInterThree)+(rightup.east |- rightup.south)$)
	  {threerep={0}{0}{0}{-2}{-1}};
\pic[anchor=east, local bounding box=rightdown] at ($(-0.4,\ecartInterThree)+(rightmid.east |- rightmid.south)$)
	  {threerep={0}{0}{0}{0}{0}};


\node[inner sep=1pt] (timeleg) at ([xshift=8pt, yshift=-8pt]rightdown.south) {t};
\draw [thick, ->] ([yshift=-8pt] leftdown.south) to (timeleg);

\draw[dotted] ([xshift = 4pt]leftdown.east |- timeleg.south) to
	([xshift = 4pt]leftdown.east |- leftup.north);

\draw[dotted] ([xshift = 4 pt]centerdown.east |- timeleg.south) to
	([xshift = 4pt]centerdown.east |- centerup.north);

\draw[element] ([yshift=3pt]leftup.north west) rectangle
				 ([yshift=-3pt,xshift=3pt]rightdown.east |- timeleg.south);
\node[elementtitle,anchor=east,above=-1pt of leftup] (serverLeg) {\vphantom{pt}server};


\def\ecartGuys{0.25}

\pic[local bounding box=leftguy, anchor=south] at (4.5, -2.59) {leftVR=1.5};
\pic[local bounding box=frontguy, right=\ecartGuys cm of leftguy.south east]  {frontVR=1.5};
\pic[local bounding box=rightguy, right=\ecartGuys cm of frontguy.south east]  {rightVR=1.5};

\pic[local bounding box=leftfov, below=10 pt of leftguy.south] {fov={0.8}{0}};
\pic[local bounding box=frontfov, below=10 pt of frontguy.south] {fov={0.8}{1}};
\pic[local bounding box=rightfov, below=10 pt of rightguy.south] {fov={0.8}{2}};


\node[font=\footnotesize, inner sep=1pt] (highpoint) at ([yshift=-3pt]leftfov.west |- leftup.north) {bw};
\draw[thick,->] ([yshift=10pt]leftfov.west |- leftguy.north) to (highpoint);
\node[font=\footnotesize, inner sep=1pt] (rightpoint) at ([yshift=10pt]rightfov.east |- rightguy.north) {t};
\draw[thick,->] ([yshift=10pt]leftfov.west |- leftguy.north) to (rightpoint);

\draw[ultra thick] plot [smooth] coordinates {([yshift=35pt]leftfov.west |- leftguy.north)
		([yshift=50pt]frontfov.west |- leftguy.north)
		([yshift=25pt]rightfov.west |- leftguy.north)
		([yshift=40pt]rightfov.east |- leftguy.north)};

\draw[dotted] ([xshift =0.04cm]leftfov.south east) to
	([xshift =0.04cm]leftfov.east |- highpoint);

\draw[dotted] ([xshift =0.04cm]frontfov.south east) to
	([xshift =0.04cm]frontfov.east |- highpoint);

\pic[local bounding box=leftBW] at ([yshift=18pt] leftguy.north){emptyEquirec={\sizeLow}};
\node[font=\scriptsize] at (leftBW) {\vphantom{h}low};
\pic[local bounding box=frontBW] at ([yshift=18pt] frontguy.north){emptyEquirec={\sizeBig}};
\node[font=\scriptsize] at (frontBW) {\vphantom{h}high};
\pic[local bounding box=rightBW] at ([yshift=18pt] rightguy.north){emptyEquirec={\sizeLow}};
\node[font=\scriptsize] at (rightBW) {\vphantom{h}low};

\draw[element] ([yshift=3pt]highpoint.west |- leftup.north) rectangle
				 ([yshift=-3pt,xshift=3pt]rightfov.south east);
\node[elementtitle, anchor=east] at (rightguy |- serverLeg) {\vphantom{pt}client};


\coordinate (client) at (highpoint.west);
\coordinate (server) at ([xshift=3pt]rightdown.east);

\tikzset{
	proto/.style={
     	-latex, thick
     }
}

\tikzset{
	protoleg/.style={
		sloped,
		inner sep=1pt,
		font=\tiny,
		above
     }
}

\tikzset{
	pics/reqresp/.style n args={2}{
		code={
			\draw[proto] (0,0) to
				node[protoleg, midway] {#1} ([yshift=-10 pt]server |- 0,0);
			\draw[proto] ([yshift=-11 pt] server |- 0,0) to
				node[draw=none,midway,matrix] {#2} ([yshift=-21pt] 0,0);
			}
	}
}

\pic[local bounding box=mpd] at (client) {reqresp={connect}
	{\node[font=\tiny,inner sep=1pt,thin,draw=gray,fill=white] at (0,0) {mpd};\\}
	};
\pic[local bounding box=oneseg] at ([yshift=-3pt]client|-mpd.south) {reqresp=
	{s$_1$:QER$_2$\,lo}
	{\pic at (0,0) {equirec={\sizeLow}{-2}{-1}};\\}
	};
\pic[local bounding box=twoseg] at ([yshift=-3pt]client|-oneseg.south)
	{reqresp={s$_2$:QER$_3$\,hi}
	{\pic at (0,0) {equirec={\sizeBig}{0}{0}};\\}
	};
\pic[local bounding box=thirdseg] at ([yshift=-3pt]client|-twoseg.south)
	{reqresp={s$_3$:QER$_1$\,lo}
	{\pic at (0,0) {equirec={\sizeLow}{2}{1}};\\}
	};


\end{tikzpicture}
   \caption{Viewport-adaptive streaming system: the server offers \num{6} representations (\num{3} \acp{QER} at \num{2} bit-rates). The streaming session lasts for three segments. The client head moves from left to right, the available bandwidth varies. For each segment, the client requests a representation that matches both the \FoV{} and the network throughput.}
   \label{fig:newdelivery}
\end{figure}
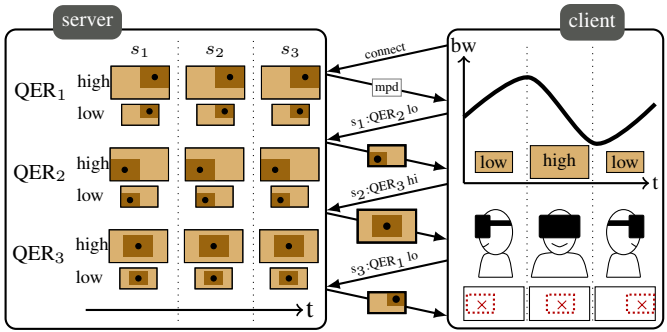

\parag{Server}The server takes as an input a $360$-degree video in
equirectangular format and transforms each frame into a desired
geometrical layout. Then, it creates $n$ different video
versions, each with a different \ac{QER} and encoded in $k$ different
bit-rates (see Figure~\ref{fig:newdelivery}). The server
splits all such encoded videos into segments, which are classified in
$n\!\times\!k$ representations (based on their respective bit-rate and
\ac{QER}), enabling clients to regularly switch from one
representation to another. The video quality around the
\ac{QEC} is the highest, while the remaining part is encoded at lower
quality.

\parag{Client}Over time the viewer moves the head and the
available bandwidth changes. Current \acp{HMD} record changes
in head orientation through rotation around three perpendicular axes,
denoted by \emph{pitch}, \emph{yaw}, and \emph{roll}.
Head movements modify the \FoV{} center, requiring a new viewport
to be displayed. State-of-the-art \acp{HMD} can perform the
extraction~\cite{fovhmds}. The client periodically sends a request
to the server for a new segment in the representation that
matches both the new \FoV{} center and the available throughput.

\parag{Adaptation algorithm}Similarly to \ac{DASH}, the client runs
an adaptation algorithm to select the video representation. It first
selects the \ac{QER} of the video based on the \FoV{} center and
the \acp{QEC} of the available \acp{QER}. This is an important addition to
the \ac{DASH} bit-rate adaptation logic, since the \ac{QER} determines
the quality of the video that is delivered and displayed to the user.
After the \ac{QER} selection, the client chooses the video
representation characterized by this \ac{QER} and whose bit-rate fits
with the expected throughput for the next $x$ seconds (\textit{i.e.},
$x$ being the segment length). The server replies
with the requested video representation, from which the
client extracts the viewport, displaying it on the \ac{HMD}, as
shown in Figure~\ref{fig:newdelivery}.

Rate-adaptive streaming systems are based on the assumption that
the selected representation will match the network
conditions for the next $x$ seconds. Rate adaptation algorithms
are
developed~\cite{tian,liu} to reduce the mismatch between the
requested bit-rate and the throughput. In our proposal, the adaptation algorithm should also ensure
that the \FoV{} centers will be as close as possible to the \ac{QEC} of the chosen \ac{QER} during
the $x$ next seconds.
In this paper, we implement
a simple algorithm for \ac{QEC} selection: we select the \ac{QEC} that
has the smallest orthodromic distance\footnote{The shortest distance
between two points on the surface of a sphere, measured along the
surface of the sphere. Its measure is proportional to the radius
of the sphere; we refer to ``distance unit'' to denote the
radius size.} to the \FoV{} center at the time the client runs
the adaptation algorithm. Similarly as for bit-rate adaptation, we
expect new viewport-adaptive algorithms to be developed in
the future to better predict the head movement and select the
\ac{QEC} accordingly. In their recent paper,~\citet{allthings} have
made a first study where they show that a simple linear regression algorithm
enables an accurate prediction of head movements for short segment size.

\parag{Video segment length}A video segment length determines how
often requests can be sent to the server. It typically ranges from
\SIrange{1}{10}{\second}. Short segments enables quick
adaptation to head movement and bandwidth changes, but it increases
the overall number of segments and results in larger manifest files.
Shorter segments also increase the network overhead due to
frequent requests, as well as the network delay because of the round
trip time for establishing a TCP connection.
Longer segments improve the encoding efficiency and quality relative to
shorter ones, however they reduce the flexibility to adapt the video
stream to changes. We discuss segment length and head movement in
Section~\ref{subsec:segmentLength} based on a dataset.

\begin{lstlisting} [float, language=xml, frame=single, backgroundcolor=\color{white},lineskip={-1pt}, caption=Extensions of MPD file,captionpos=b, label=mpdChanges]
<?xml version="1.0"?>
<MPD>
  <Representation id="1" qec="90,60" bandwidth="9876" width="1920" height="1080" frameRate="30">
   <EssentialProperty schemeIdUri="urn:mpeg:dash:vrd:2017" value="0,0">
   <SegmentList timescale="1000" duration="2000">
   ...
  </Representation>
 </AdaptationSet>
</MPD>
</xml>
\end{lstlisting}

\parag{Extending the \ac{MPD} file} To implement the proposed
\FoV{}-adaptive video streaming, we extended a \ac{DASH} \ac{MPD}
file with new information, as illustrated in Listing~\ref{mpdChanges}.
Each representation contains the \texttt{coordinates} of its \ac{QEC}
in degrees, besides the parameters that are
already defined in the standard~\cite{iso_iec}.
Those coordinates are the two angles of the spherical coordinates of the \ac{QEC}, ranging respectively from \SIrange{0}{360} degrees and from \SIrange{-90}{90} degrees. All representations from the same adaptation set should have the same reference coordinate system.
The \texttt{@schemeIdUri} is used to indicate some extra information on the video such as the video source id and the projection type. The projection type is used by the client to determine if he knows how to extract viewports from this layout.

\newcommand\testbitrateBudget{6}
\newcommand\testbitrateBudgetPercentage{\SI{75}{\percent}}
\section{System Settings}
\label{sec:settings}

The preparation of $360$-degree videos for viewport-adaptive streaming
relies on multiple parameters. We distinguish between global
parameters (the number of \acp{QER}, the number of
representations, the segment length and the geometric layout)
and local (\emph{per representation}) parameters (the target bit-rate,
the number of different qualities in a representation, the quality
arrangement of different faces of a geometric layout). We \GS{will not
be}{} comprehensive regarding the selection of all these
parameters here. Some of them require a deeper study related to signal
processing, while others depend on business
considerations and infrastructure investment. In this paper, we
restrict our attention to three key questions: What is the best
geometric layout to support quality-differentiated $360$-degree video?
What is the best segment length to support head movements, while
maintaining low management overhead? What is the best number of
\acp{QER} $n$ to reduce the induced storage requirements, while
offering a good \ac{QoE}? To answer these three questions, we have
developed a software tool and used a dataset from a real \ac{VR} system.

\parag{Dataset}We graciously received from Jaunt, Inc a dataset
recording the head movements of real users watching $360$-degree videos.
The dataset is the same as the one used by~\citet{yu_framework_2015}. It comprises
eleven omnidirectional videos that are ten seconds long. These videos
are typical of \ac{VR} systems. The dataset contains
the head movements of eleven people who were asked to watch the videos
on a state-of-the-art \ac{HMD} (Occulus Rift DK2). The subjects were
standing and they were given the freedom to turn around, so the head
movements are of wider importance than if they were asked to watch the
video while sitting. Given the length of the video and the
experimental conditions, we believe that the head movements thus
correspond to a configuration of wide head movements, which
is the most challenging case for our viewport-adaptive
system. \citet{yu_framework_2015} studied the most frequent head
positions of users. We are interested here in head movements
during the length of a segment.

\parag{Software}We have developed our own tool to manipulate the main
concepts of viewport-adaptive streaming. Since the code is publicly
available,\footnote{\url{https://github.com/xmar/360Transformations}}
the software can be used to make further studies and to
develop real systems. The main features include:
\begin{itemize}
  \item \emph{Projection from a spherical video onto any of the four geometric
layouts and vice versa}. The spherical video is the pivot format from
which it is possible to project to any layout.
Our tool rotates the
video so that the \ac{QEC} is always at the same position on
the $2$D layout.

  \item \emph{Adjusting the video quality for each
geometric face of any layout}. For each face, we set the resolution in
number of pixels and the target encoding bit-rate.

  \item \emph{Viewport extraction for any
\FoV{} center on the sphere}. It includes the decoding, rescaling
and ``projection'' of each face of the input video to extract the
viewport. This tool support extraction of \FoV{} that overlap on
multiple faces with different resolution and bit-rate target.
\end{itemize}

\subsection{Geometric Layout}

We report now the experiment of measuring the video quality of
viewports, extracted from $360$-degree videos
projected onto various geometric layouts and with various face quality
arrangements. We used two reference videos.
\begin{itemize}
   \item \emph{The original equirectangular video at full quality}:\footnote{\url{https://youtu.be/yarcdW91djQ}}
	We extract viewports at $1080$p resolution from this $4$K
   equirectangular video, which represents the reference (original)
   video 
   used to assess the objective video quality.

   \item \emph{The same equirectangular video re-encoded at a target bit-rate}.
   It is what a regular delivery system would deliver for the same bit-rate budget (here \SI{\testbitrateBudget}{\mega
   bps} being \testbitrateBudgetPercentage{} of the original video bit-rate).
   We re-encoded the original full-quality video with \ac{HEVC}
   by specifying this bit-rate target. We call it \emph{uniEqui} to
   state that, in this video, the quality is uniform.
\end{itemize}

The performance of the layout can be studied with
regards to two aspects: $(i)$ \emph{the best viewport quality}, which
is the quality of the extracted viewport when the \FoV{} center and
the \ac{QEC} perfectly matches, $(ii)$ and the \emph{sensitivity to
head movements}, which is the degradation of the viewport quality when
the distance between the \FoV{} center and the \ac{QEC} increases.
To examine both aspects, we select one \ac{QEC} on the spherical video.
We chose one orthodromic distance $d$ that will vary from \numrange{0}{\pi}.
We extract a ten seconds long viewport video, at distance $d$
from the \ac{QEC}, at the same spherical position on the original
equirectangular video and on the tested video. We used two objective
video quality metrics to measure the quality of the extracted viewport
compared to the original full quality viewport:
\ac{MS-SSIM}~\cite{wang2003multiscale} and
\ac{PSNR}.
\ac{MS-SSIM} compares image by image the structural
similarity between this video and the reference video. The \ac{PSNR} measures the average error of pixel intensities
between this video and the reference video.
The \ac{MS-SSIM} metric is closer to human perception but is less appropriate
than the \ac{PSNR} to measure difference between two measurements.
Since we
compare several encoded versions of the \emph{same} viewport against the
original, these well-known tools provide a fair performance evaluation of viewport distortion.
We perform multiple quality assessment (typically forty) at the same distance $d$ but at different positions and average the result.

\begin{figure}
\pgfplotscreateplotcyclelist{My color list}{%
    {color1,solid, very thick},%
    {color2,densely dashed, very thick},%
    {color3,densely dotted, very thick},%
    {color4,dash pattern=on 4pt off 1pt on 4pt off 4pt, very thick}%
}

\pgfplotsset{every axis legend/.append style={
        at={(-0.03,0.97)},
anchor=south west,
draw=none,
fill=none,
legend columns=4,
column sep=6pt,
font = \scriptsize,
/tikz/every odd column/.append style={column sep=0cm},
}}
\pgfplotsset{grid style={dashed,gray}}
\pgfplotsset{minor grid style={dotted,red!20!gray}}
\pgfplotsset{major grid style={dotted,green!50!black}}

\tikzsetnextfilename{distance_quality}
\begin{tikzpicture}
    \begin{axis}[
            ylabel={MS-SSIM},
            xlabel={Orthodromic distance},
            width=1.05\linewidth,
            height=0.5\linewidth,
            cycle list name={My color list},
            legend cell align=left,
            xmin = 0,
            xmax = 3.14,
            ymax = 1,
            ymajorgrids,
            yminorgrids,
        ]
         \pgfplotsextra{\begin{scope}[on layer=axis background]
                \draw[draw=color5] (axis cs:0,0.955777) -- (axis cs:3.14,0.955777);
                \node[rounded corners, fill=color5,
                			font=\tiny, inner sep=2pt,
                			anchor=west] at (axis cs:0,0.955777) {\textit{\vphantom{lj}uniEqui}};
            \end{scope}
        }

        \addplot+ table [x=distance, y=qualityEquirectangularTiledLower]{distanceQuality.csv};
        \addplot+ table [x=distance, y=qualityCubeMapLower]{distanceQuality.csv};
        \addplot+ table [x=distance, y=qualityPyramidLower]{distanceQuality.csv};
        \addplot+ table [x=distance, y=qualityRhombicDodecaLower]{distanceQuality.csv};
        \legend{Equirec,CubeMap,Pyramid,Dodeca}

    \end{axis}
\end{tikzpicture}
       \caption{Average \acs{MS-SSIM} depending on the distance to the \acs{QEC} for the four geometric layouts. Global bit-rate budget \SI{\testbitrateBudget}{\mega bps}}
    \label{fig:dist_quality}
\end{figure}
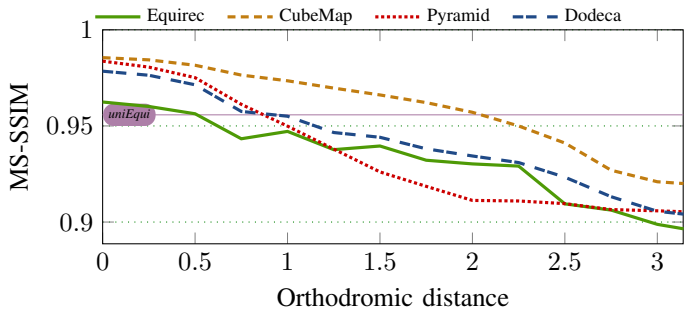

We represent in Figure~\ref{fig:dist_quality} the video quality
(measured by \acs{MS-SSIM}) of the viewport that is extracted from our
quality-differentiated layouts (equirectangular panorama with
$8\!\times\! 8$ tiles, cube map, pyramid, and dodecahedron). We also
represent by a thin horizontal line the video quality of the same
viewports extracted from the \textit{uniEqui} layout (it does
not depend on the distance since the quality is uniform). For each
geometric layout, we have tested numerous quality arrangements with
respect to the overall bit-rate budget. We selected here the ``best"
arrangement for each layout. For the cube map,
the
\ac{QEC} is located at the center of a face. This face is set at full
quality (same bit-rate target as the same portion of the original
video), and the other faces at \SI{25}{\percent} of the
full quality target.

The projection on a cube map appears to be the best choice for the
\ac{VR} provider. The quality of the viewport when the \ac{QEC} and
the \FoV{} center matches ($d=0$) is above \num{0.98}, which
corresponds to imperceptible distortion relative to the full quality
video. For all layouts, the quality decreases when the distance $d$
increases but the quality for the cube map layout is always the
highest. Note that the pyramid projection (the layout chosen
by Facebook~\cite{facebook}) is especially sensitive to head movements.
The viewport extracted from a cube map projection has a better quality
than that extracted from the \emph{uniEqui} for \FoV{} center for up to
\num{2} units from the \ac{QEC} while the other layouts viewports increase a video quality for only \num{1} unit of the \ac{QEC}.
We study next the interplay between this distance, the segment length and the number of \acp{QEC}.

\subsection{Segment Length}
\label{subsec:segmentLength}

The segment length is a key aspect of viewport-adaptive streaming.
Long segments are easier to manage and better for video encoding, but
short segments enable fast re-synchronisation to head movement. With
respect to Figure~\ref{fig:dist_quality}, the segment length
should be chosen such that the distance between the \FoV{} center and
the \ac{QEC} are rarely higher than \num{1.5}~distance units.

\begin{figure}
\centering
\pgfplotscreateplotcyclelist{My color list}{%
    {color1,solid, very thick},%
    {color2,densely dashed, very thick},%
    {color3,densely dotted, very thick},%
    {color4,dash pattern=on 4pt off 1pt on 4pt off 4pt, very thick}%
}

\pgfplotsset{every axis legend/.append style={
        at={(0.05,0.97)},
anchor=south west,
draw=none,
fill=none,
legend columns=4,
column sep=15pt,
/tikz/every odd column/.append style={column sep=0cm},
}}

\pgfplotsset{grid style={dashed,gray}}
\pgfplotsset{minor grid style={dotted,red!20!gray}}
\pgfplotsset{major grid style={dotted,gray!50!black}}

\tikzsetnextfilename{cdf_head_position_dataset}
\begin{tikzpicture}
    \begin{axis}[
            ylabel={CDF},
            xlabel={Orthodromic Distance to the Initial Head Position},
            width=1.05\linewidth,
            height=0.5\linewidth,
    	    xtick={0,1, 2, 3},
            ytick={0,0.2,0.4,...,1},
            enlarge x limits=0.02,
            cycle list name={My color list},
            legend cell align=left,
            ymin = 0,
            ymax = 1,
            ymajorgrids,
            xmajorgrids,
            y filter/.code={\pgfmathparse{#1/100}\pgfmathresult},
        ]

			\foreach \window in {1s, 2s, 3s, 5s}{
            \addplot+ table [x =\window, y = cdf]{global.csv};
        }
        	\legend{1\,s, 2\,s, 3\,s, 5\,s}

    \end{axis}
\end{tikzpicture}
\caption{CDF of the time spent at distance $d$ from the head position on the beginning of the
segment, for various segment lengths.}\label{cdf-dataset}
\end{figure}
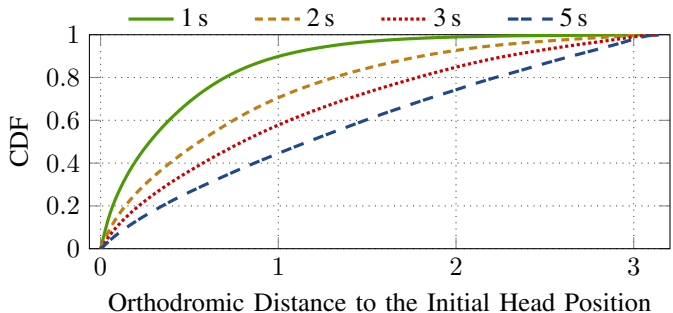

Given the dataset, we show the distribution of head movements for
various segment lengths in Figure~\ref{cdf-dataset}. For each video
and person watching it, we set timestamps that correspond to the
starting time of a video segment, \textit{i.e.,} the time at which the
users select a \ac{QEC}. Then, we measure the orthodromic distance
between this initial head position and every \FoV{} center during
the next $x$ seconds, where $x$ is the segment length. In
Figure~\ref{cdf-dataset}, we show the \ac{CDF} of the time spent at a
distance $d$ from the initial head position.
A point at $(1.5,0.6)$
means that, on average, users spend \SI{60}{\percent} of their time
with a \FoV{} center at less than \num{1.5}~distance units from the
\FoV{} center on the beginning of the segment.

Our main observation is that viewport-adaptive streaming requires
short segment lengths, typically smaller than \SI{3}{\second}. Indeed,
for a segment length of \SI{5}{\second}, users spend on average half of
their time watching at a position that is more than
\num{1.3}~distance units away from the initial head position, which
results in a degraded video quality. A segment length of
\SI{2}{\second} appears to be a good trade-off: \SI{92}{\percent} of
users never diverged to a head position that is further than
\num{2}~distance unit away from the initial head position, and users
can experience the full video quality three quarters of the time (head distance
lesser than \num{0.7}~distance unit). Please recall that our dataset
captures a challenging experiment for our system. We can expect
narrower head movements, and thus longer possible segment lengths, for
sitting users and longer videos. Note also that these results are consistent
with the head movement prediction from~\citet{allthings}, who showed that
prediction accuracy drops for time periods greater than \SI{2}{\second}.

\subsection{Number of \acp{QER}}

The number of \acp{QER} $n$ represents another key trade-off. The more
\acp{QER} there are, the better the coverage of the spherical video
is, and thus the better the viewport quality will be due to a better
match between the \ac{QEC} and the \FoV{} center. However,
increasing the number of \acp{QER} also means increased storage and
management requirements at the server (and a longer \ac{MPD} file).

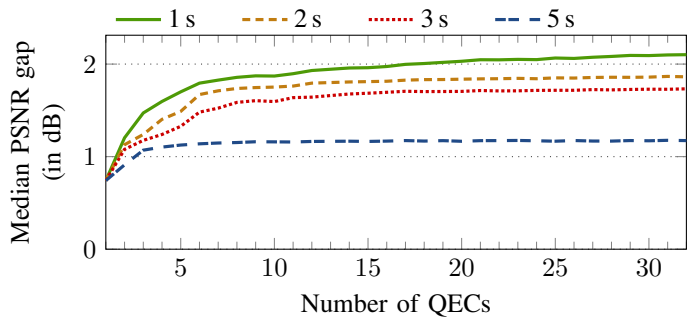
\begin{figure}
\centering
\pgfplotscreateplotcyclelist{My color list}{%
    {color1,solid, very thick},%
    {color2,densely dashed, very thick},%
    {color3,densely dotted, very thick},%
    {color4,dash pattern=on 4pt off 1pt on 4pt off 4pt, very thick},%
    {color5,dotted, very thick},%
}

\pgfplotsset{every axis legend/.append style={
        at={(0,0.97)},
anchor=south west,
draw=none,
fill=none,
legend columns=5,
column sep=15pt,
/tikz/every odd column/.append style={column sep=0cm},
}}

\pgfplotsset{grid style={dashed,gray}}
\pgfplotsset{minor grid style={dotted,red!20!gray}}
\pgfplotsset{major grid style={dotted,gray!50!black}}

\tikzsetnextfilename{qec_nb_to_qoe}
\begin{tikzpicture}
    \begin{axis}[
            ylabel style={align=center}, ylabel={Median PSNR gap\\(in dB)},
            xlabel={Number of QECs},
            width=1.05\linewidth,
            height=0.5\linewidth,
    	    xtick={0,5,...,50},
           minor x tick num={4},
            cycle list name={My color list},
            legend cell align=left,
            xmin = 1,
            ymin = 0,
            xmax = 32,
            ymajorgrids,
        ]
        \foreach  \window in {1, 2, 3, 5}{
         \addplot+ table [x =nbQec, y =medQoe]{qoeForWindow\window s.csv};
        }
        \legend{1\,s, 2\,s, 3\,s, 5\,s}

    \end{axis}
\end{tikzpicture}
\caption{Median \acs{PSNR} gap between the viewports of the cube map layout and the \textit{uniEqui} depending
on the number of \acp{QER}.
Bit-rate: \SI{\testbitrateBudget}{\mega bps}}
\label{fig:QEC}
\end{figure}

We represent in Figure~\ref{fig:QEC} the median \ac{PSNR} difference
between the viewport extracted from the cube map layout and the
same viewport extracted from the \emph{uniEqui} layout with the same
overall bit-rate budget. To modify the number of \acp{QER}, we set a
number $n$, then we determined the position of the $n$ \acp{QEC}
using the Thomson positioning problem~\cite{rakhmanov1994electrons}.
For each head position in the dataset, we computed the distance
between the \FoV{} center and the \ac{QEC} that was chosen at the
beginning of the segment and we computed the viewport quality
accordingly.

The best number of \acp{QER} in this configuration is between
\numlist{5;7}. The gains that are obtained for higher number of
\acp{QER} are not significant enough to justify the induced storage
requirements (in particular not $30$~\acp{QER} as in the Facebook
system~\cite{facebook}). Having multiple  \acp{QER} provides higher
quality gains for short segments, due to the better re-synchronization
between the \acp{QER} and the \FoV{} centers. Note 
that a
significant part of these gains stems from the cube map layout. 

\section{Conclusion}
\label{sec:conclusion}

We have introduced in this paper viewport-adaptive streaming for
navigable $360$-degree videos. Our system aims at offering both
interactive highs-quality service to \ac{HMD} users with low management
for \ac{VR} providers.
We studied the main system
settings of our framework, and validated its relevance.
We emphasize that, with current encoding techniques, the cube
map projection for two seconds segment length and six
\acp{QER} offers the best performance. This
paper opens various research questions:
(I) New adaptation algorithms should be studied for viewport navigation,
especially based on head movement prediction techniques using \emph{saliency maps} (probability of
presence), extracted from the feedback of
previous viewers~\cite{han2014spatial}. \citet{allthings} have recently made a first attempt in this
direction.
(II) New video encoding methods should be
developed to perform quality-differentiated encoding for
large-resolution videos. Especially, methods that allow for
intra-prediction and motion vector prediction across \emph{different} quality
areas. The recent work from~\citet{vishyArxiv} is a first step.
(III) Specific studies for \emph{live} \ac{VR} streaming
and interactively-generated $360$-degree videos should be performed,
because the different representations can hardly be all
generated on the fly.


%
%
%
%
%
%

\bibliographystyle{abbrvnat}
\bibliography{biblio}

\end{document}